\documentstyle[twocolumn,aps]{revtex}

\input epsf

\begin{document}
\title{Rotational Feshbach Resonances in Ultracold Molecular Collisions}
\author{J. L. Bohn$^1$, A. V. Avdeenkov$^1$, and M. P. Deskevich$^2$}
\address{$^1$JILA and Department of Physics,
University of Colorado, Boulder, CO 80309 \\
$^2$Department of Chemistry and Biochemistry, University of Colorado, Boulder, CO
80309}
\date{\today}
\maketitle

\begin{abstract}
In collisions at ultralow temperatures, molecules will possess Feshbach 
resonances, foreign to ultracold atoms, whose virtual excited states 
consist of rotations of the molecules.  We estimate the mean
spacing and mean widths of these resonant states, exploiting the fact
the molecular collisions at low energy display chaotic motion.
As examples, we consider the experimentally relevant molecules O$_2$, OH,
and PbO.  The density of s-wave resonant states for these species
is quite high, implying that a large number of narrow resonant states
will exist.  
\end{abstract}


\narrowtext

Scattering resonances are of great importance in ultracold collisions.
Feshbach resonances occur when the energy of a pair of free atoms (or molecules)
is nearly degenerate with that of a quasi-bound state of the pair.  The
quasi-bound state is characterized by the promotion of one or both atoms
to an excited internal state, for example an excited hyperfine state
\cite{Stwalley}.   External magnetic fields can then shift these states
in or out of resonance, giving the experimenter direct control over
interparticle interactions \cite{Tiesinga}.  In this way, Bose-Einstein
condensates of $^{85}$Rb have been made stable or unstable on command, leading to
novel many-body effects \cite{Weiman}.  It is moreover predicted that this
kind of control will be useful in preparing degenerate Fermi gases in
``resonant superfluid'' states with any desired interaction strength 
\cite{Holland,Timmermans}.
Magnetic field Feshbach resonances have now been observed in the alkali
atoms $^{23}$Na \cite{Ketterle}, $^{85}$Rb \cite{Heinzen,Roberts}, $^{133}$Cs
\cite{Chu}, and $^{40}$K \cite{Jin}.  While not uncommon, these resonances
are far from ubiquitous in ultracold alkali atoms.  This is largely due to 
the relatively small number of hyperfine states available to form resonances.

Ultracold molecules, by contrast, offer a far greater number
of resonances than alkali atoms, because rotational excitations can also
contribute to resonant states.  In contrast to the two hyperfine states 
in an alkali atom, a molecule can possess
a large number of energetically available rotational states.  Considering
that ``typical'' rotational energy splittings are of order 1-10 K, while
the well depths of intermolecular potential energy surfaces (PES's) can be
hundreds or thousands of K, it is clear that tens of rotational states may contribute,
including their degeneracies arising from magnetic quantum numbers.

Identifying the number and properties of these resonances is 
important for understanding the behavior of a molecular gas reduced to
extremely low temperatures.   These resonances may be useful as tools
for pinning down the details of multidimensional PES's.  They may also
imply resonant control over cold molecular collisions, including, possibly,
control over chemical reactions \cite{Bala}.  In addition,
the scattering dynamics is chaotic, implying that ultracold resonances
will form a new laboratory for studying quantum chaos \cite{Qchaos}.
In this Letter we  estimate the most basic properties of these
resonances, namely, how many we can expect, and what their widths might be.
We are interested here in molecules produced in their ro-vibrational
ground states by buffer-gas cooling \cite{Doyle} or Stark slowing
\cite{Meijer} techniques.  However, vibrationally excited cold molecules
produced by photoassociation of cold atoms will
also exhibit rich resonant dynamics \cite{Forrey}.

Classical chaos at low energies has been studied extensively in 
molecular scattering problems \cite{Atkins}.  Of greatest relevance to 
our present purposes are multiple collision events: it is possible 
that a collision deposits sufficient energy in internal molecular degrees 
of freedom (e.g., rotations and vibrations) that there is not enough
translational energy left to allow the molecules to separate.
The molecules may therefore collide many times before finally 
shedding enough energy to separate.  Multiple collisions are 
increasingly likely as the collision energy is lowered, and in fact 
are useful in quantifying the onset of chaos in this regime
\cite{Pattard}.

\epsfxsize = 3in
\begin{figure}
\epsfbox{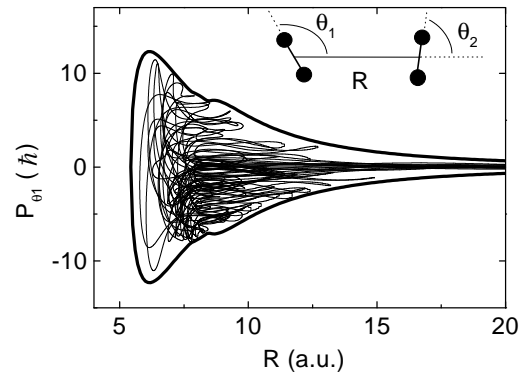}
\vskip .5in
\caption{A classical trajectory for O$_2$-O$_2$ scattering at a collision
energy 1 mK.  See text for details.}
\end{figure}

Our classical calculations verify that multiple-collision resonances
persist at ultracold temperatures.  We consider scattering of $^{17}$O$_2$
molecules, using the singlet rigid rotor potential energy surface of Ref. 
\cite{Avoird}.  At the collision energies considered, vibrational
excitation is not energetically allowed.  To simplify the calculation
we constrain all the coordinates to fixed values except the intermolecular
distance $R$ and the orientations of the molecules in the scattering plane, as
shown in the inset to Figure 1.   The main part of the figure shows
a slice through a sample phase space trajectory for collision energy 1 mK.
The axes are $R$ and the angular momentum $P_{\theta1}$
(in units of $\hbar$) of one of the molecules.  As a guide, the 
heavy solid line indicates the allowed region of phase space.
For the molecules to separate at this energy, the greatest 
angular momentum either molecule could have is $0.027 \hbar$.
Instead, the molecules can exchange several $\hbar$ of angular momentum
in each collision, and the trajectory eventually fills the allowed 
phase space.

\epsfxsize = 3in
\begin{figure}
\epsfbox{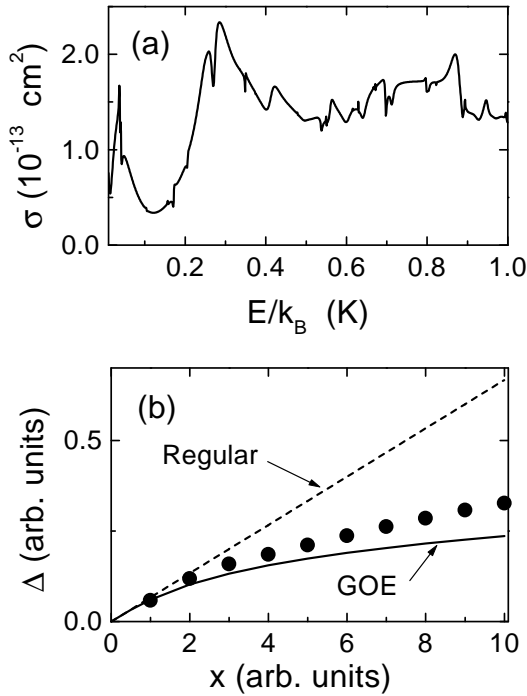}
\vskip .5in
\caption{Evidence of chaos in quantum mechanical scattering of
oxygen molecules.  (a) shows the complete elastic scattering
cross section, computed as in Ref. [19].  In (b) the Dyson
rigidity function (points) is computed for the energy spacings of
the 33 resonances in (a).  Our results are much more closely
related to the results of a gaussian orthogonal ensemble (GOE)
approximation to a chaotic system than to a nonchaotic regular
system.}
\end{figure}

Chaotic motion in the classical realm leaves its signature in
the quantum mechanical spectrum as well.  We have computed the quantum-
mechanical elastic scattering cross section for the $|N=0,J=1,M_J=1 
\rangle$ fine-structure state of $^{17}$O$_2$ [Fig. 2(a)], following the model
described in Ref. \cite{Avdeenkov}.  As in Ref. \cite{Avdeenkov}, 
we have neglected the nuclear spin of the $^{17}$O atoms, to emphasize 
the effect of from rotational motion.   This
model includes rotational levels up to $N=2$, and the partial
waves $L=0,2,...14$, and finds 33 resonances in this energy range.
Treating these levels as bound states of the (O$_2$)$_2$ collision
complex, we expect to find evidence of quantum chaos in statistical measures of 
the energy level distributions.  A glance at the spectrum in Fig. 2a)
suggests that there is no clustering of levels, but rather that they
are roughly evenly spaced.  This ``rigidity'' of level spacings 
is characteristic of the quantum mechanical
spectrum of a system that displays classical chaos, and is usually
quantified in terms of the spectral rigidity function $\Delta (x)$
of Dyson and Mehta \cite{Dyson}.  If the number of states with energy 
less than or equal to $E$ is denoted $N(E)$, then $\Delta (x)$
represents the root-mean-squared deviation of $N(E)$ from a
straight line over an energy range $x$, where $x$ is measured in units of the
mean level spacing.  Figure 2(b) shows this function, computed using the
resonances from Fig 2(a).  A spectrum consisting of uncorrelated energy levels  would
yield a rigidity $\Delta (x) = x/15$, as shown with the dashed line.  
A spectrum generated in the Gaussian orthogonal ensemble (GOE) approximation,
representing a chaotic system, yields instead the solid line.
The $\Delta (x)$ computed from the spectrum in Fig. 2(a) more closely
resembles the GOE result, suggesting that this system is
chaotic.  A similar analysis has been applied to the eigenphase shifts
in reactive scattering processes \cite{shifts}.

Given that the classical trajectories can fill phase space (Fig.1),
each energetically allowed state is equally likely to be populated
during a resonant collision.  This observation has been exploited in the theory of
unimolecular chemical reactions,  which considers the breakup of a
large molecule under the influence of random collisions with other molecules
in a solution \cite{unimol}.  We use the same idea here to estimate
the total number of resonant states available in a given energy range.
To this end we separate the molecule-molecule Hamiltonian into three 
independent terms:
\begin{equation}
{\hat H} = {\hat H}_1 + {\hat H}_2 + {\hat H}_{\rm int},
\end{equation}
where ${\hat H}_1$ and ${\hat H}_2$ represent the rotational fine structure
of each individual molecule,  and ${\hat H}_{\rm int}$ is an effective
intermolecular potential that depends only on $R$.
 Each Hamiltonian is represented by a set of
independent energy eigenvalues $E_1$, $E_2$, and $E_{\rm int}$.
During a resonant collision any partition of the
total energy $E = E_1 + E_2 + E_{\rm int}$ is equally likely, provided that
angular momentum is conserved.  Estimating the number of states $N(E)$
at or below energy $E$ is then a simple counting exercise.  The density of
states, $\rho(E) = dN/dE$, is also easily computed.

To illustrate that this is a reasonable estimate of the density of states,
we compare the results to those of a fully quantum mechanical calculation
for $^{17}$O$_2$ - $^{17}$O$_2$ scattering \cite{Avdeenkov}.  
Table I tabulates, for the quantum mechanical calculation, the number of
resonances in each partial wave $L$ in an energy range 1K above the incident
threshold.  Each row in the table represents a separate calculation, 
characterized by the maximum number of molecular rotation states ($N_{\rm max}$)
and partial wave states ($L_{\rm max}$) included in the basis set.

Table II reports the same information, but
this time estimated from the statistical model. Statistical fluctuations in the
density of states function $\rho(E)$ in the relevant energy range
correspond to uncertainties of $\sim 50$ \% in the numbers given.
Within this uncertainty the agreement with the quantum mechanical
calculation in Table I is quite good, giving us further confidence that the 
statistical estimate is reasonable for this purpose.

A clear advantage of the statistical model over the full calculation is that 
it can easily be extrapolated to include {\it all} allowed  values of
$N$ and $L$ consistent with conservation of energy and angular
momentum, as well as Bose symmetry.  The result
is given by the final row in Table II, for small values of the incident 
partial wave.  Considering all partial waves yields a total number of
resonances in this 1K interval of order $\sim 10^3$.  However, in ultracold
collisions we are limited to low values of $L$, typically $L=0$ alone,
barring occasional shape resonances with higher values of $L$.  We therefore 
conclude from Table II that the mean density of s-wave rotational resonances 
in O$_2$ is on the order of 10 K$^{-1}$.  For comparison, alkali atoms 
typically exhibit on the order of 1 s-wave Feshbach resonance per 1 K energy 
interval.  Thus rotational Feshbach resonances are slightly more numerous 
than atomic hyperfine Feshbach resonances, but not strikingly so, at least for
nonpolar molecules.
(In practice the numerous hyperfine levels in $^{17}$O$_2$ will play a 
role also, but we are concerned here  strictly with  rotation.)

This simple statistical theory can also estimate the widths of these resonances.
Again borrowing from the theory of unimolecular dissociation, the
resonance width may be approximated in the 
Rice-Ramsperger-Kassel-Marcus (RRKM) approximation as
\cite{unimol}
\begin{equation}
{\bar \Gamma } = {W(E^+) \over 2 \pi {\bar \rho} },
\end{equation}
where ${\bar \rho}$ is the mean density of states we have been considering,
and $W(E^+)$ is the number of states that permit the molecules to
actually escape to infinite separation.  For the low collision energies
considered here, both molecules must return to their ground state to separate,
so that $W(E^+)=1$.  In the quantum mechanical calculation in Fig. 2(a),
there are 33 resonances in a 1 Kelvin interval, implying a mean width
in this approximation of ${\bar \Gamma} \approx 5$ mK.  For comparison,
the arithmetic mean of the quantum mechanical resonance widths is 8 mK.
Note that the quantum mechanical widths span several orders of magnitude; 
their standard deviation is not given by the RRKM result \cite{note}.

The situation is dramatically different for polar molecules, whose
long range dipole-dipole interaction potential, proportional to $1/R^3$,
holds many more intermolecular bound states near threshold.
To illustrate this point we consider two examples: OH,
which is suitable for Stark slowing \cite{Meijer}; and the $a(1)$ state
of PbO, which is a leading candidate in which to measure the electric dipole
moment of the electron \cite{DeMille,Egorov}.  Both molecules possess,
in addition to a dipole moment, a pair of closely-spaced states of opposite
parity:  a $\Lambda$ doublet of splitting $\Delta = 0.081$ K in OH, 
and an $\Omega$-doubling of splitting $\Delta =260$ $\mu$K in PbO.
Intermolecular potential curves correlating to these thresholds will
hold a large number of dipole-bound resonant states.  Details of
the fine structure of OH is given in \cite{Coxon}, and that of PbO 
in \cite{Martin}.

For the OH-OH interaction an approximate PES exists \cite{Kuhn}.
We include the long-range part of this interaction, which has
a minimum at $R \approx 6$ a.u. that arises from hydrogen bonding
forces.  For the sake of our statistical model, we approximate 
this surface with a single curve by fixing the two OH molecules in
their linear geometry.  The intermolecular bound state energies 
$E_{\rm int}$ are then simply the bound states of this potential,
for each of the partial waves considered.  We ignore for our present
purposes an even deeper part of the PES corresponding to chemical binding
of hydrogen peroxide, H$_2$O$_2$.
For PbO-PbO the PES is entirely unknown; however, since the bound states of interest
depend primarily on the long-range potential, we estimate the PbO-PbO
PES by multiplying that of OH-OH by the appropriate ratio of squared
dipole moments, $[d_{\rm PbO}/d_{\rm OH}]^2 \approx 7.7$.  Table III tabulates
the number of s-wave resonances to be expected for each species.
These numbers refer to the important energy range $2 \Delta$ above 
threshold, where the number of resonances is greatest.  Expressed as 
a mean density-of-states over this range, the densities are 240 
K$^{-1}$ for OH and $2 \times 10^5$ K$^{-1}$ for PbO, 
significantly larger than for the nonpolar O$_2$.

In general, the resonances shift when an external electric field
is applied, so that one after the other can be brought into resonance
with the essentially zero-energy molecules.  Given a mean density of
states $\approx 10^2$ K$^{-1}$ for OH, 
and given the OH dipole moment (=1.67 D), the mean spacing
of these resonances as a function of field would be of order $\sim 250$ V/cm.
This means that, in contrast to the magnetic field resonances in alkali
atoms, there will be many available resonances in ultracold polar
molecule collisions.

We conclude by sketching the implications of this work for the theory
of ultracold collisions.  As is well-known, ultracold collision theory
requires a certain level of experimental input to fine-tune the scattering 
Hamiltonian.  It is not yet clear how much or what kind of information
is required to perform this fine-tuning for molecules \cite{Hutson}.  
Knowing roughly how many resonances to expect can greatly simplify theory, by 
providing an estimate of how many parameters are strictly necessary
to reproduce the data.  With this information theory could then turn around 
and predict how these resonances evolve with applied external electric and
magnetic fields.

A second, equally important implication concerns experiments at temperatures
where $kT$ is large compared to the mean spacing between resonances.
In this case individual resonances are not resolved, and only the overall 
modulation of the spectrum matters.  It is then likely that 
semiclassical methods \cite{Delos,Granger} should prove insightful.

This work was supported by the NSF and by the OSEP program at the 
University of Colorado.  We acknowledge an allocation of
parallel computer time by NPACI, and useful discussions with
D. Nesbitt.

\begin{table}
\caption{Number of resonances in a 1K energy interval above threshold for 
cold collisions of $^{17}$O$_2$
molecules in their $J=M_J=+1$ state, neglecting nuclear spin.  These results
are determined from a quantum mechanical calculation.  Various levels 
of completeness of this calculation are specified by the maximum value 
of rotational quantum number $N_{\rm max}$ and maximum partial wave 
$L_{\rm max}$.  For each such calculation, the table gives the number 
of resonances found separately in each incident partial wave $L$.}
\label{table1}
\begin{tabular}{cccccccc}
$N_{\rm max}$ & $L_{\rm max}$ & 
$L=0$ & $L=2$ & $L=4$ & $L=6$ & $L=8$ & $L=10$ \\
\tableline
2 & 2 & 1 & 4 & $-$ & $-$ & $-$ & $-$ \\
2 & 4 & 2 & 7 & 8 & $-$ & $-$ & $-$ \\
2 & 6 & 2 & 7 & 9 & 8 & $-$ & $-$ \\
2 & 8 & 2 & 10 & 14 & 9 & 10 & $-$ \\
2 & 10 & 3 & 10 & 14 & 11 & 11 & 6 \\
4 & 4 & 3 & 11 & 15 & $-$ & $-$ & $-$ \\
\end{tabular}
\end{table}

\begin{table}
\caption{Number of resonances for $^{17}$O$_2$ cold collisions, 
as in Table I.  Here, however, the 
resonances are estimated using the statistical counting model described 
in the text.  In addition, the final row extrapolates to the limit where
$N_{\rm max}=8$ and $L_{\rm max}=36$, thus including all possible
energetically allowed states of the $(^{17}$O$_2)_2$ dimer.}
\label{table2}
\begin{tabular}{cccccccc}
$N_{\rm max}$ & $L_{\rm max}$ &
$L=0$ & $L=2$ & $L=4$ & $L=6$ & $L=8$ & $L=10$ \\
\tableline
2 & 2 & 2 & 5 & $-$ & $-$ & $-$ & $-$ \\
2 & 4 & 3 & 8 & 11 & $-$ & $-$ & $-$ \\
2 & 6 & 3 & 10 & 15 & 11 & $-$ & $-$ \\
2 & 8 & 3 & 10 & 16 & 14 & 9 & $-$ \\
2 & 10 & 3 & 10 & 16 & 16 & 12 & 9 \\
4 & 4 & 5 & 15 & 24 & $-$ & $-$ & $-$ \\
\tableline
All & All & 14 & 45 & 78 & 83 & 85 & 84 \\
\end{tabular}
\end{table}

\begin{table}
\caption{Summary of s-wave resonances for two polar molecules. 
$\rho_{\Delta}$ is the average density of states consistent with
s-wave initial conditions, over an 
energy range $2 \Delta$ above threshold, where $\Delta$ refers to the 
$\Lambda$ ($\Omega$)-doublet of OH (PbO). Using the RRKM result,
this density translates into the mean energy width ${\bar \Gamma}_{\Delta}$.}
\label{table3}
\begin{tabular}{cccc}
Molecule & $\Delta$ (K) & $\rho _{\Delta}$ (K$^{-1}$) & ${\bar \Gamma}_{\Delta}$
($\mu$K)\\ 
\tableline
OH  & 0.081 & 240 & 660\\
PbO & $2.6 \times 10^{-4}$ & $1.9 \times 10^5$ & $5 $\\
\end{tabular}
\end{table}


\begin{references}
\bibitem[*]{byline} email address: bohn@murphy.colorado.edu
\bibitem{Stwalley} W. C. Stwalley, Phys. Rev. Lett. {\bf 37}, 1628 (1976).
\bibitem{Tiesinga} E. Tiesinga, B. J. Verhaar, and H. T. C. Stoof,
Phys. Rev. A {\bf 47}, 4114 (1993).
\bibitem{Weiman} J. L. Roberts {\it et al.}, Phys. Rev. Lett. {\bf 86}, 4211 (2001).
\bibitem{Holland} M. Holland, S. J. J. M. Kokkelmans, M. L. Chiofalo,
and R. Walser, Phys. Rev. Lett. {\bf 87}, 120406 (2001). 
\bibitem{Timmermans} E. Timmermans, K. Furuya, P. W. Milloni, and 
A. K. Kerman, Phys. Lett, A {\bf 285}, 228 (2001).
\bibitem{Ketterle} S. Inouye {\it et al.}, Nature {\bf 392}, 151 (1998).
\bibitem{Heinzen} Ph. Courteille {\it et al.}, Phys. Rev. Lett. {\bf 81}, 69 (1998).
\bibitem{Roberts} J. L. Roberts {\it et al.}, Phys. Rev. Lett. {\bf 81}, 5109 (1998).
\bibitem{Chu} V. Vuleti\'{c}, A. J. Kerman, C. Chin, and S. Chu, Phys. Rev. Lett.
{\bf 82}, 1406 (1999).
\bibitem{Jin} T. Loftus {\it et al.}, Phys. Rev. Lett. {\bf 88}, 173201 (2002).
\bibitem{Bala} That chemical reactions may occur at ultralow temperatures
is suggested by N. Balakrishnan and A. Dalgarno, Chem. Phys. Lett. {\bf 341},
652 (2001).
\bibitem{Qchaos} M. C. Gutzwiller, {\it Chaos in Classical and Quantum Mechanics} 
(New York, Springer-Verlag, 1990).
\bibitem{Doyle} J. D. Weinstein {\it et al.}, Nature {\bf 395}, 148 (1998).
\bibitem{Meijer} H. L. Bethlem {\it et al.}, to appear in Phys. Rev.  A.
\bibitem{Forrey} R. C. Forrey {\it et al.}, Phys. Rev. Lett. {\bf 82}, 2657 (1999).
\bibitem{Atkins} K. M. Atkins and J. M. Hutson, J. Chem. Phys.
{\bf 103}, 9218 (1995), and references therein.
\bibitem{Avoird} P. E. S. Wormer and A. van der Avoird, J. Chem. Phys. 
{\bf 81}, 1929 (1984).
\bibitem{Pattard} T. Pattard and J. M. Rost, Chem. Phys. Lett. {\bf 291}, 360
(1998).
\bibitem{Avdeenkov} A. V. Avdeenkov and J. L. Bohn, Phys. Rev. A {\bf 64},
052703 (2001).
\bibitem{Dyson} F. J. Dyson and M. L. Mehta, J. Math. Phys. {\bf 4},
701 (1963).
\bibitem{shifts} P. Honvault and J.-M. Launay, Chem. Phys. Lett.  {\bf 329}, 
233 (2000). 
\bibitem{unimol} R. G. Gilbert and S. C. Smith, {\it Theory of Unimolecular
and Recombination Reactions} (Blackwell Scientific, 1990).
\bibitem{note} In addition, resonance widths can display signautres of
quantum chaos, but the limited statistics of this sample prevent us from
drawing any conclusions in this regard.
\bibitem{Kuhn} B. Kuhn {\it et al.}, J. Chem. Phys. {\bf 111}, 2565 (1999).
\bibitem{DeMille} D. DeMille {\it et al.}, Phys. Rev. A {\bf 61}, 052507 (2000).
\bibitem{Egorov} D. Egorov {\it et al.}, Phys. Rev. A {\bf 63}, 030501 (2001).
\bibitem{Coxon} J. A. Coxon, Can. J. Phys. {\bf 58}, 933 (1980).
\bibitem{Martin} F. Martin {\it et al.}, Spectrochim. Acta {\bf 44A}, 889 (1988).
\bibitem{Hutson} M. Muewly and J. M. Hutson, J. Chem. Phys. {\bf 110},
8338 (1999).
\bibitem{Delos} M. L. Du and J. B. Delos, Phys. Rev. A {\bf 38},
1896 (1988); {\it ibid.}, {\bf 38}, 1913 (1998).
\bibitem{Granger} B. E. Granger and C. H. Greene, Phys. Rev. A {\bf 62},
012511 (2000).

\end{references}
\end{document}